\documentclass[a4paper,11pt]{article}
\usepackage{multirow}
\usepackage{color}
\usepackage{latexsym}
\usepackage{amssymb}
\usepackage{amsfonts}
\usepackage{amsmath}
\usepackage{amsthm}
\usepackage{indentfirst}
\usepackage{mathrsfs}
\usepackage{verbatim}

\usepackage{tikz}
\usetikzlibrary{patterns}

\usepackage[T1]{fontenc}

\theoremstyle{remark}

\usepackage[latin1]{inputenc}
\usepackage{diagbox}

\usepackage{fullpage}

\author{Antonio Bernini\thanks{Dipartimento di Matematica e Informatica ``U. Dini", University of Firenze, Firenze, Italy\quad \texttt{\{antonio.bernini, stefano.bilotta, elisa.pergola\}@unifi.it}}\and Stefano Bilotta$^\star$\and Elisa Pergola$^\star$}
	
\title{An Identity for Catalan Numbers via Restricted Dyck Paths}

\date{}

\begin{document}
	
	\maketitle
	
	\begin{abstract}
		Catalan numbers and their interpretations in terms of Dyck paths are widely used in different topics of applied mathematics and computer science. Here, we consider a general approach for constrained Dyck paths.
		In particular, we study Dyck paths of height at most $h$ with the additional restriction of having no $k-1$ consecutive valleys at height $h-1$. We give a combinatorial description of this class of paths and derive enumeration formulas using classical techniques for counting constrained lattice paths.
		As a consequence of this analysis, we obtain an identity involving Catalan numbers which, to the best of the authors' knowledge, does not appear in the existing literature. This identity arises naturally from the combinatorial interpretation and provides a new relation among families of Dyck paths with height and local structural constraints.
	\end{abstract}
	
	Keywords: Catalan numbers, combinatorial identity, pattern avoidance.
	
	\section{Introduction}
	The Catalan numbers form a classical sequence in enumerative combinatorics and arise in a wide variety of counting problems across applied mathematics and computer science. The $n$-Catalan number is defined by
	$$C_n=\frac{1}{n+1}\binom{2n}{n},$$ 
	and counts numerous families of combinatorial objects, including balanced parenthesis expressions, rooted ordered trees, binary trees, non-crossing partitions, and many more. A comprehensive survey of Catalan numbers and their many interpretations can be found in \cite{Stanley}.
	Among most notable objects enumerated by the Catalan numbers, there are Dyck paths. Dyck paths are the canonical representative of the class of Catalan objects since they have been widely used in several applications (a formal definition of Dyck paths is presented in the next Section). In computer science, Dyck paths provide a natural combinatorial model for well-formed structures and languages, giving fundamental results in theory of automata, parsing and syntax analysis \cite{K}. Moreover, the bijections between Dyck paths, trees, and stack-based processes make them a central tool in the analysis of recursive algorithms and data structures, where Catalan numbers frequently appear in both exact and asymptotic complexity analyses.
	
	In coding theory, Dyck paths are used to model bifix-free codes, synchronization codes, and other classes of constrained codes. Here, we only recall the involvement of Dyck paths in Gray codes \cite{BBP2}, non-overlapping codes \cite{bi}, cryptography \cite{SAB} and partially ordered structures \cite{BCFS}.
	
	Dyck path enumeration has also received much attention in recent decades. An interesting paper dealing with this matter is the one by Deutsch \cite{D} where the author enumerates Dyck paths according to various parameters. 
	
	A subclass of these
	paths has been considered thanks to the simple
	behaviour of the recursive relations describing them and the rational nature of the associated  generating functions. 
	More precisely, 
	the generating function associated with Dyck paths is algebraic, and it is rational when the paths are bounded \cite{BM,BMP}, for example with respect to the height. Kallipoliti et al.\ \cite{KST} consider Dyck paths of height less or equal to a precise value. Moreover, in the same paper, the analysis of some characteristics of Dyck paths avoiding valleys at specified height is considered. Taking into account the same line of research, the authors in \cite{Dh2} consider Dyck paths of height equal or less than $h$ and with no valley at height $h-1$, while in \cite{BBP6} they consider Dyck paths of height at most $2$ and $3$ avoiding a generic number of consecutive valleys at height $1$ and $2$, respectively. In both papers they obtain interesting Catalan identities.
	
	In the present paper, we address the open problem proposed in \cite{Dh2}, which concerns a generalization that does not depend on the number of avoided valleys. As results, we solve the above mentioned open problem and we introduce a new approach to counting Catalan numbers under specific constraints. To the best of  our knowledge, this relation is new in the literature and leads to a novel recurrence formulation for Catalan numbers and their applications. We also point out that the obtained identity differs from the ones of \cite{Dh2,BBP6} since the recurrence relation for the $n$-th Catalan number has constant coefficients under particular hypothesis.

	The paper structure is the following. In Section \ref{Prelim} some preliminaries on Dyck paths are presented. Section \ref{Section3} is devoted to summarize the results in \cite{Dh2} and generalize them to an arbitrary number of consecutive valleys, providing a generating algorithm for considered Dyck paths. Some enumerative results are presented in Section \ref{enum}, where the related formal proof is also detailed. In Section \ref{catalan} we present a new intriguing relation involving Catalan numbers which is a direct consequence of the results obtained from the constrained Dyck paths.
	Finally, we conclude the paper proposing some further developments on the present topics.
	
	\section{Notation}\label{Prelim}
	A \emph{Dyck path} is a lattice path in the discrete plane $\mathbb Z^2$ from $(0,0)$ to $(2n,0)$ with up and down steps in $\{(1,1),(1,-1)\}$, never crossing the $x$-axis. The number of up steps in every prefix of a Dyck path is greater than or equal to the number of down steps, and the total number of steps (the \emph{length} of the path) is $2n$. We denote the set of Dyck paths of length $2n$ (or equivalently semilength $n$) by $\mathcal D_n$.
	A Dyck path can be codified by a string over the alphabet $\{U,D\}$, where $U$ and $D$ replace the up and down steps, respectively. The empty Dyck path is denoted by $\varepsilon$.
	
	The \emph{height} of a Dyck path $P$ is the maximum ordinate reached by one of its steps.
	A \emph{valley} of $P$ is an occurrence of the substring $DU$, while a \emph{peak} is an occurrence of the substring $UD$. The height of a valley (peak) is the ordinate reached by $D$ ($U$). With the notation $U^\ell \ (D^\ell)$ we intend a consecutive sequence of $U \ (D)$ steps of length $\ell$.   
	
	We denote by $\mathcal D_n^{(h,k)}$ the set of Dyck paths having semilength $n$ and height at most $h$, and avoiding $k-1$ consecutive valleys at height $h-1$. Moreover, we set
	$$ \mathcal D^{(h,k)}=\displaystyle\bigcup_{n\geq 0} \mathcal D_n^{(h,k)}.
	$$
	The cardinality of $\mathcal D_n^{(h,k)}$ is indicated by $D_n^{(h,k)}$. Finally,
	the set $\mathcal D_n$ of unrestricted Dyck paths having semilength $n\geq 0$ is enumerated by the $n$-Catalan number
	$C_n=\frac{1}{n+1}\binom{2n}{n}.$
	
	The set $\mathcal D_n^{(h,2)}$ represents the set of Dyck paths with height at most $h$ and without valley at height $h-1$. The algorithmic construction approach and the related enumeration of the set $\mathcal D_n^{(h,2)}$ are studied in a recent paper \cite{Dh2}. Moreover, a bijection with 312-avoiding permutations having some restriction on their left to right maxima is also provided in \cite{Dh2}. In the present work we generalize the construction and enumeration of $\mathcal D_n^{(h,2)}$ to $\mathcal D_n^{(h,k)}$ with $k \geq 2$. We also provide its generating function according to the semilength and a novel identity involving Catalan numbers. 
	
	\section{Generation}\label{Section3}
	According to \cite{Dh2} the set $\mathcal D^{(h,2)}$ can be exhaustively generated by means of an ECO operator \cite{BDPP1} which allows to construct all the paths of a certain size $n+1$ starting from the ones of size $n$. The generating algorithm is synthetically described by a succession rule, where the role of each label $(k)$ of a $n$-size path $P$ is the number of the $k$ paths generated by $p$ having size $n+1$. See \cite{Dh2} for details.
	
	In particular, the generating algorithm for $\mathcal D^{(h,2)}$, for $h\geq 3$, can be described by the following succession rule:
	\begin{equation}\label{rule_h2}
		\Omega_{h,2}:
		\begin{cases}
			(1)&\\
			(1)\leadsto(2)&\\
			(l)\leadsto(2)(3)\cdots(l)(l+1),& \text{for $2\leq l<h$}\\
			(h)\leadsto(2)(3)\cdots(h-1)^2(h).&
		\end{cases}
	\end{equation}
	
	Now, we are going to describe a new ECO operator $\theta$ leading to the general succession rule $\Omega_{h,k}$ describing the construction of $\mathcal D^{(h,k)}$, for $h\geq4$ and $k\geq 3$ (note that for $h=2$ and $h=3$, the results can be found in \cite{BBP6}).

	The idea is to consider certain \emph{sites} in each path $P \in \mathcal D_{n}^{(h,k)}$ where the insertion of the factor $\mathbf{UD}$ is allowed, so as to generate every path in $\mathcal D_{n+1}^{(h,k)}$ exactly once. As $n$ ranges from $0$ to infinity, the generation of $\mathcal D^{(h,k)}$ is obtained. In the ECO methodology framework \cite{BDPP1}, these sites are called \emph{active sites}.
	%without repetitions
	%from $P$ (so that the sites are called \emph{active sites}).
	
	In general, given a path $P$ of semilength $n$, the insertion of the 
	factor $\mathbf{UD}$ before or after each letter of $P$ gives rise to a 
	path $Q$ of semilength $n+1$. Therefore, the \emph{sites} consist of 
	the points between two consecutive letters of $P$, together with the 
	points preceding the first letter and following the last letter. If $P$ 
	is read as a lattice path in the discrete plane, it is not difficult to 
	realize that the \emph{sites} are the initial points of each up or down 
	step of $P$, together with the final point of the last down step.
	%An operator $\phi$ ($\phi: \mathcal D_n^{(h,k)} \to \mathcal
	% D_{n+1}^{(h,k)}$) acts by inserting $\mathbf{UD}$ exactly at the
	% sites.
	For example, if $P=UD$, then $P$ has three \emph{sites}, and the insertion of $\mathbf{UD}$ in all the sites produces the paths $\{\mathbf{UD}UD, U\mathbf{UD}D, UD\mathbf{UD}\}$.
	
	According to the definition of ECO operator \cite{BDPP1} we have to characterize an operator $\theta: 
	\mathcal D_n^{(h,k)}\to \mathcal D_{n+1}^{(h,k)}$ such that the two following properties are satisfied:
	\begin{enumerate}
		\item given $P,Q\in \mathcal D_n^{(h,k)}$, then $\theta(P)\cap\theta(Q)=\emptyset$;
		
		\item for each $Q'\in\mathcal D_{n+1}^{(h,k)}$ there exist $Q\in \mathcal D_n^{(h,k)}$ such that $Q'\in\theta (Q)$.  
	\end{enumerate}
	
	A path $P=U_1 U_2 \cdots U_t D P' \in \mathcal D_{n}^{(h,k)}$ (where $1\leq t \leq h$) starts with at most $h$ up steps $U$, and $P'$ is a Dyck suffix of length $n-t-1$ avoiding $k-1$ consecutive valleys. The operator $\theta$ works by inserting the factor $\mathbf{UD}$ at selected sites (the \emph{active sites}) of the first run of $P$, ensuring that the two properties above are fulfilled. We have the two following cases:
	
	\begin{itemize}
		\item[i)] if $P=U_1 U_2 \cdots U_{t-1}U_t D P' \in \mathcal D_n^{(h,k)}$, with $t=h$, then $P$ has one of the following factorizations highlighting the number of consecutive valleys occurring at height $h-1$ in the prefix before $P'$:
		\begin{itemize}
			\item[$\bullet$] $P_0=U^hD^2P'$ ($0$ valleys at height $h-1$);
			\item[$\bullet$] $P_1=U^h(DU)D^2P'$ ($1$ valleys at height $h-1$);
			\item[$\bullet$] $P_2=U^h(DU)^2D^2P'$ ($2$ valleys at height $h-1$);\\
			\vdots
			\item[$\bullet$] $P_{k-3}=U^h(DU)^{k-3}D^2P'$ ($k-3$ valleys at height $h-1$);
			\item[$\bullet$] $P_{k-2}=U^h(DU)^{k-2}D^2P'$ ($k-2$ valleys at height $h-1$).
			
		\end{itemize}
		
		In each kind of these paths $P$ the factor $\mathbf{UD}$ can be inserted at the starting point of the initial up steps of $P$, having height $0,1,2,\cdots,h-2$ (the height of a point is intuitively its ordinate). We claim (without proof for the moment) that these points form a subset of the active sites of $P$.
		%These points are surely \emph{active sites} in $P$: the insertion
		%of $UD$ in one of them produces a path in $\mathcal 
		%D_{n+1}^{(h,k)}$.
		
		In the case of the site at height $h-1$, the 
		insertion of $\mathbf{UD}$ is allowed only in the paths $P_{\ell}=U^h(DU)^\ell 
		D^2P'$ with $\ell=0,1,\cdots,k-3$ (clearly, the index $\ell$ denotes the number of consecutive valleys having height $k-1$ in the prefix before $P'$). Indeed, the insertion of $\mathbf{UD}$ in the site having height $h-1$ in $P_{k-2}$ generates $Q=U^{h-1}\mathbf{UD}U(DU)^{k-2}P'=U^h(DU)^{k-1}P'$, so that $Q \notin \mathcal D_n^{(h,k)}$. Equivalently, the site at height $h-1$ cannot serve as an active site for $P_{\ell}$ when $\ell=k-2$, while otherwise it can.
		
		Summarizing, the insertion of $\mathbf{UD}$ in the identified sites gives rise to:
		\begin{align*}
			\theta(P_\ell)=\{&\mathbf{UD} U^h (DU)^{\ell}D^2 P',\\
			&U\mathbf{UD} U^{h-1} (DU)^{\ell}D^2 P'=U^2 D U^{h-1} (DU)^{\ell}D^2P',\\
			& \cdots\\
			&U^{h-1}\mathbf{UD} U(DU)^{\ell}D^2 P'=U^h (DU)^{\ell+1}D^2 P'\ \},
		\end{align*}
		for $\ell=0,1,2,\cdots,k-3$, and the cardinality of $\theta(P_{\ell})$ is $h$. While, for $\ell=k-2$ we have:  
		\begin{align*}
			\theta(P_{k-2})=\{&\mathbf{UD} U^h (DU)^{k-2}D^2 P',\\
			&U\mathbf{UD} U^{h-1} (DU)^{k-2}D^2 P'=U^2 D U^{h-1} (DU)^{k-2}D^2P',\\
			& \cdots\\
			&U^{h-2}\mathbf{UD} U^2(DU)^{k-2}D^2 P'=U^{h-1} DU^2(DU)^{k-2}D^2 P'\ \},
		\end{align*}
		
		and the cardinality of $\theta(P_{k-2})$ is $h-1$.
		
		It is evident that $\mathbf{UD}$ can not be inserted without taking 
		account the number $\ell$ of consecutive valleys after the $h$ 
		initial up steps, so that, in the case $t=h$, we have to 
		distinguish between the possible values of $\ell$. To this aim, we 
		label the Dyck paths 
		$$P_{\ell}=U^h(DU)^\ell D^2P'$$ 
		for	$\ell=0,1,\cdots,k-3$, respectively by $(h_0),(h_1),\cdots, (h_{k-3})$. 
		Clearly, the path $P_{k-2}$ has label $(h-1)$.
		
		\item[ii)] if $P=U_1 U_2 \cdots U_{t-1}U_t D P' \in \mathcal 
		D_n^{(h,k)}$, with $1\leq t < h $, all the sites before the first 
		$t+1$ steps can be used for the insertion of $\mathbf{UD}$, 
		generating:
		\begin{align*}
			\theta(P)=\{&\mathbf{UD} U_1 U_2 \cdots U_{t-1}U_t D P',\\
			&U_1\mathbf{UD} U_2 \cdots U_{t-1}U_t D P',\\
			& \cdots\\
			&U_1 U_2 \cdots U_{t-1} \mathbf{UD} U_t D  P',\\
			&U_1 U_2 \cdots U_{t-1} U_t \mathbf{UD} D P'\ \}
		\end{align*}
		Hence, the path $P$ has label $(t+1)$, and the cardinality of $\theta(P)$ is $t+1$.
	\end{itemize}

	The restriction of $\theta$ to the sites identified as active satisfies the properties required for $\theta$ to be an ECO operator.
	While the first of the two is straightforward to verify, for the second one we observe that, if $Q'\in\mathcal D_{n+1}^{(h,k)}$ it suffices to remove the first peak from the left to obtain a path $Q\in\mathcal D_n^{(h,k)}$ such that $Q'\in\theta(Q)$.
	
	Having defined the $\theta$ operator, the labels of each path in $\theta(P)$ can be easily retrieved, leading to the following succession rule:
	\begin{equation}\label{rule_hk}
		\Omega_{h,k}:
		\begin{cases}
			(1)&\\
			(1)\leadsto(2)&\\
			(2)\leadsto(2)(3)&\\
			\vdots&\\
			(h-1)\leadsto(2)(3)\cdots(h-1)(h)&\\
			(h)\leadsto(2)(3)\cdots(h-1)(h)(h_0)&\\
			(h_0)\leadsto(2)(3)\cdots(h-1)(h)(h_1)&\\
			(h_1)\leadsto(2)(3)\cdots(h-1)(h)(h_2)&\\
			\vdots&\\
			(h_{k-3})\leadsto(2)(3)\cdots(h-1)^2(h)\ ,
		\end{cases} 
	\end{equation}
	where the label $(1)$ (the axiom of the succession rule) is related to the empty path $\varepsilon$ which produces, by means of $\theta$, the path $UD$ having label $(2)$.
	
	\section{Generating function} \label{enum}
	
	The aim of this section is to find the generating function of the paths in $\mathcal D^{(h,k)}$, according to their semilength $n$.
	
	Let $P \in \mathcal D^{(h,k)}$, we denote by $|P|$ its semilength and by $F(P)$ the number of paths of semilength $|P|+1$ in $\theta(P)$. Moreover, we assign to each path $P$ a label $\ell(P)$. Then, from succession rule (\ref{rule_hk}), 
	$$\ell(P) \in \{(1),(2),\ldots,(h),(h_0),(h_1),\ldots, (h_{k-3})\}$$ and 
	$$F(P) \in \{1,2,\ldots,h\}.$$
	Let
	
	\begin{align*}
		f(x,y,y_0,y_1,\ldots,y_{k-3})=&\sum_{\substack{P \in \mathcal D^{(h,k)} \\ \ell(P)=(1),\ldots,(h)}} x^{|P|} y^{F(P)} + \sum_{\substack{P \in \mathcal D^{(h,k)} \\ \ell(P)=(h_0)}} x^{|P|} y_0^{h} + \sum_{\substack{P \in \mathcal D^{(h,k)} \\ \ell(P)=(h_1)}} x^{|P|} y_1^{h}+ \cdots \\
		&\\
		& \cdots + \sum_{\substack{P \in \mathcal D^{(h,k)} \\ \ell(P)=(h_{j})}} x^{|P|} y_{j}^{h} + \cdots + \sum_{\substack{P \in \mathcal D^{(h,k)} \\ \ell(P)=(h_{k-3})}} x^{|P|} y_{k-3}^{h}
	\end{align*}
	be the $k$-variate generating function of $\mathcal D^{(h,k)}$, where $x$ keeps track of the semilength of $P$ and $y,y_0,y_1, \ldots, y_{k-3}$ keep track of $F(P)$. Note that this definition takes into account the fact that, for paths $P$ with $\ell(P)=(h_0),(h_1),\ldots,(h_{k-3})$, one has $F(P)=h$.  
	
	\bigskip
	We define:
	%\begin{align*}
	%	&F_1(x,y,y_0,y_1,\ldots,y_{k-3})= \sum_{\substack{P \in \mathcal D^{(h,k)} \\ \ell(P)=(1)}} x^{|P|} y\\
	%	&\\
	%	&F_2(x,y,y_0,y_1,\ldots,y_{k-3})= \sum_{\substack{P \in \mathcal 
			%		D^{(h,k)} \\ \ell(P)=(2)}} x^{|P|} y^2\\
	%	&\vdots\phantom{aaaaaaaaaaaaaaa}\\
	%	&\\
	%	&F_h(x,y,y_0,y_1,\ldots,y_{k-3})= \sum_{\substack{P \in \mathcal   
			%		D^{(h,k)} \\ \ell(P)=(h)}} x^{|P|} y^h\\
	%	&\\
	%	&T_0(x,y,y_0,y_1,\ldots,y_{k-3})= \sum_{\substack{P \in \mathcal 
			%		D^{(h,k)} \\ \ell(P)=(h_0)}} x^{|P|} y_0^h\\
	%	&\\
	%	&T_1(x,y,y_0,y_1,\ldots,y_{k-3})= \sum_{\substack{P \in \mathcal 
			%		D^{(h,k)} \\ \ell(P)=(h_1)}} x^{|P|} y_1^h\\
	%	&\vdots\phantom{aaaaaaaaaaaaaaa}\\
	%	&\\
	%	&T_{k-3}(x,y,y_0,y_1,\ldots,y_{k-3})= \sum_{\substack{P \in 
			%		\mathcal D^{(h,k)} \\ \ell(P)=(h_{k-3})}} x^{|P|} y_{k-3}^h \ .
	%\end{align*}
	$$
	F_r(x,y,y_0,y_1,\ldots,y_{k-3})= \sum_{\substack{P \in \mathcal D^{(h,k)} \\ \ell(P)=(r)}} x^{|P|} y^r \mbox{\ \ for } r=1,2,\ldots,h
	$$
	and
	$$
	T_s(x,y,y_0,y_1,\ldots,y_{k-3})= \sum_{\substack{P \in \mathcal 
			D^{(h,k)} \\ \ell(P)=(h_s)}} x^{|P|} y_s^h \mbox{\ \ for } s=0,1,\ldots,k-3.
	$$
	
	Clearly, we have:
	\begin{equation}\label{genFunk-var}
		f(x,y,y_0,y_1,\ldots,y_{k-3})=\sum_{i=1}^h F_{i}(x,y,y_0,y_1,\ldots,y_{k-3}) + \sum_{i=0}^{k-3} T_{i}(x,y,y_0,y_1,\ldots,y_{k-3}) \ .
	\end{equation}

	From (\ref{rule_hk}) we deduce:
	\begin{itemize}
		\item $F_1(x,y,y_0,y_1,\ldots,y_{k-3}) = y$, since the only path with $\ell(P) = (1)$ is the empty Dyck path $\varepsilon$.
		
		\item For $i=2,3,\ldots,h-2$, we have: 
		%	\[
		%	\begin{aligned}
			%		F_j(x,y,y_0,y_1,\ldots,y_{k-3}) = 
			%		&\sum_{\substack{P \in \mathcal{D}^{(h,k)} \\ \ell(P) = (j-1)}} x^{|P|+1} y^j 
			%		+ \sum_{\substack{P \in \mathcal{D}^{(h,k)} \\ \ell(P) = (j)}} x^{|P|+1} y^j 
			%		+ \ldots \\
			%		&+ \sum_{\substack{P \in \mathcal{D}^{(h,k)} \\ \ell(P) = (h)}} x^{|P|+1} y^j 
			%		+ \sum_{\substack{P \in \mathcal{D}^{(h,k)} \\ \ell(P) = (h_0)}} x^{|P|+1} y^j 
			%		+ \ldots \\
			%		&+ \sum_{\substack{P \in \mathcal{D}^{(h,k)} \\ \ell(P) = (h_{k-3})}} x^{|P|+1} y^j = \\
			%		=& xy^j \displaystyle\sum_{i=j-1}^h \left(\sum_{\substack{P \in \mathcal{D}^{(h,k)} \\ \ell(P) = (i)}} x^{|P|}\right) + xy^j\sum_{\substack{P \in \mathcal{D}^{(h,k)} \\ \ell(P) = (h_0)}} x^{|P|} 
			%		+ \ldots \\
			%		&+ xy^j\sum_{\substack{P \in \mathcal{D}^{(h,k)} \\ \ell(P) = (h_{k-3})}} x^{|P|} = \\
			%		=& xy^j \displaystyle\sum_{i=j-1}^h F_i(x,1,y_0,y_1,\ldots,y_{k-3})+\\
			%		&xy^j T_0(x,y,1,y_1,\ldots,y_{k-3})+\\
			%		&xy^j T_1(x,y,y_0,1,y_2,\ldots,y_{k-3})+\\
			%		&\vdots\\
			%		&xy^j T_{k-3}(x,y,y_0,y_1,\ldots,y_{k-4},1).\\
			%	\end{aligned}
		%	\]
		%
		%	pppppp
		
		\[
		\begin{aligned}
			&F_i(x,y,y_0,y_1,\ldots,y_{k-3}) = \\
			\\
			&\sum_{\substack{P \in \mathcal{D}^{(h,k)} \\ \ell(P) = (i-1)}} x^{|P|+1} y^i 
			+ \sum_{\substack{P \in \mathcal{D}^{(h,k)} \\ \ell(P) = (i)}} x^{|P|+1} y^i 
			+ \ldots 
			+ \sum_{\substack{P \in \mathcal{D}^{(h,k)} \\ \ell(P) = (h)}} x^{|P|+1} y^i \\ \\
			&+\sum_{\substack{P \in \mathcal{D}^{(h,k)} \\ \ell(P) = (h_0)}} x^{|P|+1} y^i 
			+\sum_{\substack{P \in \mathcal{D}^{(h,k)} \\ \ell(P) = (h_1)}} x^{|P|+1} y^i
			+ \ldots 
			+ \sum_{\substack{P \in \mathcal{D}^{(h,k)} \\ \ell(P) = (h_{k-3})}} x^{|P|+1} y^i  \\ \\
			&=xy^i \displaystyle\sum_{t=i-1}^h \left(\sum_{\substack{P \in \mathcal{D}^{(h,k)} \\ \ell(P) = (t)}} x^{|P|}\right) + xy^i\sum_{\substack{P \in \mathcal{D}^{(h,k)} \\ \ell(P) = (h_0)}} x^{|P|} 
			+ \ldots 
			+ xy^i\sum_{\substack{P \in \mathcal{D}^{(h,k)} \\ \ell(P) = (h_{k-3})}} x^{|P|}  \\ \\
			&= xy^i \displaystyle\sum_{t=i-1}^h F_t(x,1,y_0,y_1,\ldots,y_{k-3})+
			xy^iT_0(x,y,1,y_1,\ldots,y_{k-3})\\\\
			&+xy^iT_1(x,y,y_0,1,y_2,\ldots,y_{k-3})+
			\ldots + xy^i T_{k-3}(x,y,y_0,y_1,\ldots,y_{k-4},1).\\
		\end{aligned}
		\]

		\item $F_{h-1}(x,y,y_0,y_1,\ldots,y_{k-3}) =$
		\[
		\begin{aligned}
			%&F_{h-1}(x,y,y_0,y_1,\ldots,y_{k-3}) = \\\\
			&\sum_{\substack{P \in \mathcal{D}^{(h,k)} \\ \ell(P) = (h-2)}} x^{|P|+1} y^{h-1} 
			+ \sum_{\substack{P \in \mathcal{D}^{(h,k)} \\ \ell(P) = (h-1)}} x^{|P|+1}+ y^{h-1}
			+ \sum_{\substack{P \in \mathcal{D}^{(h,k)} \\ \ell(P) = (h)}} x^{|P|+1} y^{h-1}\\ \\
			&+ \sum_{\substack{P \in \mathcal{D}^{(h,k)} \\ \ell(P) = (h_0)}} x^{|P|+1} y^{h-1} 
			+ \sum_{\substack{P \in \mathcal{D}^{(h,k)} \\ \ell(P) = (h_1)}} x^{|P|+1} y^{h-1}
			+ \ldots + 2\sum_{\substack{P \in \mathcal{D}^{(h,k)} \\ \ell(P) = (h_{k-3})}} x^{|P|+1} y^{h-1} \\\\
			&= xy^{h-1} \displaystyle\sum_{t=h-2}^h F_t(x,1,y_0,y_1,\ldots,y_{k-3})+xy^{h-1} T_0(x,y,1,y_1,\ldots,y_{k-3})+\\\\
			&xy^{h-1} T_1(x,y,y_0,1,y_2,\ldots,y_{k-3})+\ldots + 2 xy^{h-1} T_{k-3}(x,y,y_0,y_1,\ldots,y_{k-4},1).\\
		\end{aligned}
		\]
		Note that, from (\ref{rule_hk}), the paths $P$ with $\ell(P)=h_{k-3}$ produce two paths having label $(h-1)$. This is the reason why the last line of the formula contains the coefficient $2$.
		\item $F_{h}(x,y,y_0,y_1,\ldots,y_{k-3})=$ 
		\[
		\begin{aligned}
			%		F_{h}(x,y,y_0,y_1,\ldots,y_{k-3}) = 
			&\sum_{\substack{P \in \mathcal{D}^{(h,k)} \\ \ell(P) = (h-1)}} x^{|P|+1} y^{h} 
			+ \sum_{\substack{P \in \mathcal{D}^{(h,k)} \\ \ell(P) = (h)}} x^{|P|+1} y^{h}
			+ \sum_{\substack{P \in \mathcal{D}^{(h,k)} \\ \ell(P) = (h_0)}} x^{|P|+1} y^{h} + \ldots + \sum_{\substack{P \in \mathcal{D}^{(h,k)} \\ \ell(P) = (h_{k-3})}} x^{|P|+1} y^{h}\\\\
			& =  xy^{h} \displaystyle\sum_{t=h-1}^h F_t(x,1,y_0,y_1,\ldots,y_{k-3})+
			xy^{h} T_0(x,y,1,y_1,\ldots,y_{k-3})\\\\
			&+xy^{h} T_1(x,y,y_0,1,y_2,\ldots,y_{k-3})+
			\ldots + xy^{h} T_{k-3}(x,y,y_0,y_1,\ldots,y_{k-4},1).\\
		\end{aligned}
		\]
		\item $T_0(x,y,y_0,y_1,\ldots,y_{k-3}) =\sum_{\substack{P \in \mathcal{D}^{(h,k)} \\ \ell(P) = (h)}} x^{|P|+1} y_0^{h}=xy_0^h F_h(x,1,y_0,y_1,\ldots,y_{k-3})$.
		
		\item $T_1(x,y,y_0,y_1,\ldots,y_{k-3}) =\sum_{\substack{P \in \mathcal{D}^{(h,k)} \\ \ell(P) = (h_0)}} x^{|P|+1} y_1^{h}=xy_1^h T_0(x,y,1,y_1,\ldots,y_{k-3})$.
		
		\item $T_2(x,y,y_0,y_1,\ldots,y_{k-3}) =\sum_{\substack{P \in \mathcal{D}^{(h,k)} \\ \ell(P) = (h_1)}} x^{|P|+1} y_2^{h}=xy_2^h T_1(x,y,y_0,1,y_2,\ldots,y_{k-3})$.
		
		\hspace{.2cm} \vdots
		
		\item $T_{k-3}(x,y,y_0,y_1,\ldots,y_{k-3}) =\sum_{\substack{P \in \mathcal{D}^{(h,k)} \\ \ell(P) = (h_{k-4})}} x^{|P|+1} y_{k-3}^{h}=xy_{k-3}^h T_{k-4}(x,y,y_0,y_1,\ldots,y_{k-4},1)$.

	\end{itemize}
	
	Recalling the aim of the present section, we are interested in
	$$
	f(x,1,1,1,\ldots,1)=\sum_{i=1}^h F_{i}(x,1,1,1,\ldots,1) + \sum_{i=0}^{k-3} T_{i}(x,1,1,1,\ldots,1) \ .
	$$
	% $f(x,1,1,1,\ldots,1)$ which can be found by $F_i(x,1,1,1,\ldots,1)$ and $T_j(x,1,1,1,\ldots,1)$, 
	%with $i=1,2,\ldots,h$ and $j=0,1,\ldots,k-3$.
	For the sake of brevity, we pose 
	\begin{align*}
		F_i(x,1,1,1,\ldots,1)&=F_i(x) \mbox{ for } i=1,2,\ldots,h\ ,\\
		T_j(x,1,1,1,\ldots,1)&=T_j(x) \mbox{ for } j=0,1,\ldots,k-3, \mbox{ and }\\
		f(x,1,1,1,\ldots,1)&=f(x).
	\end{align*}
	It is easy to see that 
	$$
	T_j(x)=x^{j+1}F_h(x), \mbox{ for } j=0,1,2,\ldots,k-3\ .
	$$
	Moreover, for $i=2,\ldots,h-2$ and $i=h$, we have:
	$$
	F_i(x)=x\sum_{t=i-1}^h F_t(x) + x\sum_{j=0}^{k-3} T_j(x)= x\sum_{t=i-1}^h F_t(x)+ x^2 F_h(x)\sum_{j=0}^{k-3} x^j \ . 
	$$
	So that, the following linear system in the variables $F_i(x)$ and $T_j(x)$ for $i=1,2,\ldots,h$ and $j=0,1,\ldots,k-3$ is obtained:  
	\begin{equation}\label{F_hk}
		\begin{cases}
			F_1(x) = 1, \\[10pt]
			F_i(x) = x \left( \displaystyle\sum_{t=i-1}^{h} F_t(x) + x F_h(x) \cdot \frac{1 - x^{k-2}}{1 - x} \right), \text{for } i = 2,3,\ldots,h-2, \text{and } i=h, \\[10pt]
			\\
			F_{h-1}(x) = x \left( \displaystyle\sum_{t=h-2}^{h
			} F_t(x) + x F_h(x) \cdot \frac{1 - x^{k-2}}{1 - x} + x^{k-2} F_h(x) \right), \\[10pt]
			\\
			T_j(x)= x^{j+1}F_h(x), \text{ for } j=0,1,2,\ldots, k-3.
		\end{cases}
	\end{equation}
	Since $F_i(x)=xF_{i-1}(x) + F_{i+1}(x)$, for $i=2,3,\ldots,h-3$, system (\ref{F_hk}) boils down to:
	\begin{equation}\label{F2_hk}
		\begin{cases}
			F_1(x) = 1, \\[4pt]
			xF_1(x)-F_2(x)+F_3(x)=0, \\[4pt]
			xF_2(x)-F_3(x)+F_4(x)=0, \\[4pt]
			\vdots \\
			xF_{h-4}(x)-F_{h-3}(x)+F_{h-2}=0, \\[4pt]
			xF_{h-3}(x)-F_{h-2}(x)+F_{h-1}-x^{k-1}F_h(x)=0, \\[4pt]
			xF_{h-2}(x)-F_{h-1}(x)+(1+x^{k-1})F_h(x)=0, \\[4pt]
			xF_{h-1}(x)+(-1+x+x^2(\frac{1-x^{k-2}}{1-x}))F_h(x)=0, \\[4pt]
			T_j(x)= x^{j+1}F_h(x), \text{ for } j=0,1,2,\ldots, k-3\, 
		\end{cases}
	\end{equation}
	The first $h$ equations form an independent linear system in the variables $F_i(x)$ and by setting $$\mathbf{F}=(F_1(x),F_2(x),\dots,F_h(x))^\top \mbox{ and } 
	\mathbf{b}=(1,0,0,\ldots,0)^\top\ ,$$ this can be rewritten in matrix form as 
	$$A\,\mathbf{F}=\mathbf{b}\ ,$$
	where the matrix $A$ is
	$$
	A =
	\begin{pmatrix}
		1 & 0 & 0 & 0 & \cdots & 0 & 0 \\[6pt]
		x & -1 & 1 & 0 & \cdots & 0 & 0 \\[6pt]
		0 & x & -1 & 1 & \cdots & 0 & 0 \\[6pt]
		\vdots & \vdots & \ddots & \ddots & \ddots & \vdots & \vdots \\[6pt]
		0 & 0 & \cdots & x & -1 & 1 & -x^{k-1} \\[6pt]
		0 & 0 & \cdots & 0 & x & -1 & 1 + x^{k-1} \\[6pt]
		0 & 0 & \cdots & 0 & 0 & x & -1 + x + x^2 \tfrac{1 - x^{k-2}}{1 - x}
	\end{pmatrix}
	$$
	which exhibits an almost tridiagonal structure.

	As $T_j(x)= x^{j+1}F_h(x)$, for $j=0,1,2,\ldots, k-3$, then from (\ref{genFunk-var}) the generating function $f(x)$ of $\mathcal{D}^{(h,k)}$ according to the only semilength of the paths can be written as
	\begin{displaymath}
		f(x)
		= \sum_{i=1}^{h} F_i(x)
		+ \frac{x^{k-1} - x}{x - 1} F_h(x).
	\end{displaymath}
	
	In the next subsection we prove that the solution of $A\,\mathbf{F}=\mathbf{b}$ is given by 
	\begin{equation}\label{candidate}
		%F_{\ell}(x)= (-1)^{\binom{h \bmod 2 + \ell + 3}{2}} 
		F_i(x)= (-1)^{\binom{h \bmod 2 +i+3}{2}}
		\frac{x^{i-1}S^{(h+1-i,k)}(x)}{S^{(h,k)}(x)}, \mbox{ for } i = 1, 2, 
		\dots, h,
	\end{equation}
	where,
	\begin{align*}
		&S^{(1,k)}(x) = (x - 1),\\
		&S^{(2,k)}(x) = (2x - x^k - 1),\\
		&S^{(3,k)}(x) = (x^2 - 3x + x^{k+1} + 1),
	\end{align*}
	
	and 
	
	\begin{equation}\label{S}
		S^{(h,k)}(x)
		= \sum_{j=0}^{\left\lfloor \frac{h+1}{2} \right\rfloor}
		(-1)^{\binom{h+1}{2}-j}
		\binom{h-j+1}{j}
		x^j+
		\sum_{j=1}^{\left\lfloor \frac{h}{2} \right\rfloor}
		(-1)^{\binom{h+1}{2}-j+1}
		\binom{h-j-1}{j-1}
		x^{k+j}.
	\end{equation}
	
	\subsection{Proof}
	We prove that the expressions in (\ref{candidate}) are the solution of $A\,\mathbf{F}=\mathbf{b}$ by directly showing that they satisfy its equations. In the following we pose $$\beta=(h\bmod 2 + h + 2)\ .$$
	
	\begin{itemize}
		\item We start from the $h$-th equation of (\ref{F2_hk}). We have to prove that $$xF_{h-1}(x)+\left(-1+x+x^2\left(\frac{1-x^{k-2}}{1-x}\right)\right)F_h(x)=0\ .$$ Replacing the expressions of $F_{h-1}$ and $F_h$ derived from (\ref{candidate}), this is equivalent to prove that
		\begin{align*}
			&(-1)^{\binom{\beta}{2}} x^{h-1} S^{(2,k)}(x)-(1-x-x^2\frac{1-x^{k-2}}{1-x})(-1)^{\binom{\beta+1}{2}}x^{h-1}S^{(1,k)}(x)=\\
			&(-1)^{\binom{\beta}{2}} x^{h-1} (2x-x^k-1) -(-1)^{\binom{\beta+1}{2}}x^{h-1} (2x-x^k-1)=0.
		\end{align*}
		As $\beta$ is even, then $\binom{\beta}{2}$ and $\binom{\beta+1}{2}$ have same parity (it is easy to see that their difference is even). Therefore, both terms cancel each other, resulting in the expression being zero.
		
		%\item If $i=1$, then we have to prove 
		%$xF_{h-1}(x)+(-1+x+x^2(\frac{1-x^{k-2}}{1-x}))F_h(x)=0$. Posing $\beta=(h 
		%\bmod 2 + h + 2)$ and applying expressions in (\ref{candidate}), we have to 
		%prove that:
		%\begin{align*}
		%	&(-1)^{\binom{\beta}{2}} x^{h-1} 
		%	S^{(2,k)}(x)-(1-x-x^2\frac{1-x^{k-2}}{1-x})(-1)^{\binom{\beta+1}{2}}x^{h-1
			%	}S^{(1,k)}(x)=\\
		%	&(-1)^{\binom{\beta}{2}} x^{h-1} (2x-x^k-1) 
		%	-(-1)^{\binom{\beta+1}{2}}x^{h-1} (2x-x^k-1)=0.
		%\end{align*}
		%
		%As $\beta$ is even, then $\binom{\beta}{2}$ and $\binom{\beta+1}{2}$ have 
		%same parity (it is easy to see that their difference is even). Therefore, 
		%both terms contribute the same sign and cancel each other, resulting in the 
		%expression being zero.
		\item In order to prove the $(h-1)$-th equation of (\ref{F2_hk}), after replacing the expressions for $F_{h-2}, F_{h-1}, \mbox{ and } F_h$, we have to verify that
		\begin{equation*}
			(-1)^{\binom{\beta-1}{2}} x^{h-2} S^{(3,k)}(x)-(-1)^{\binom{\beta}{2}}x^{h-2}S^{(2,k)}(x)
			+(1+x^{k-1})(-1)^{\binom{\beta+1}{2}}x^{h-1}S^{(1,k)}(x)=0\ ,
		\end{equation*}
		which, after simplifying the factor $x^{h-2}$ and replacing the expressions of $S^{(1,k)}, S^{(2,k)}$, and $S^{(3,k)}$, is equivalent to:  
		\begin{equation*}
			(-1)^{\binom{\beta-1}{2}} (x^2-3x-x^{k+1}+1)-(-1)^{\binom{\beta}{2}} (2x-x^k-1)+ (-1)^{\binom{\beta+1}{2}} (x^2+x^{k+1}-x-x^k)=0.
		\end{equation*}
		It is easy to see that $\binom{\beta-1}{2}$ and $\binom{\beta+1}{2}$ always have different parity. 
		Moreover, since $\beta$ is even, the terms $\binom{\beta-1}{2}$ and $\binom{\beta}{2}$ also have different parity. 
		Therefore, we have either
		\[
		\binom{\beta-1}{2}\ \text{even},\quad
		\binom{\beta}{2}\ \text{odd},\ \binom{\beta+1}{2}\ \text{odd},
		\]
		or
		\[
		\binom{\beta-1}{2}\ \text{odd},\quad
		\binom{\beta}{2}\ \text{even},\ \binom{\beta+1}{2}\ \text{even}.
		\]
		In both the cases, the thesis follows.
		%\item If $i=2$, then we have to prove 
		%$xF_{h-2}(x)-F_{h-1}(x)+(1+x^{k-1})F_h(x)=0$. Posing $\beta=(h \bmod 2 + h + 
		%2)$ and applying expressions in (\ref{candidate}), we have to prove that:
		%\begin{equation*}
		%	(-1)^{\binom{\beta-1}{2}} x^{h-2} 
		%	S^{(3,k)}(x)-(-1)^{\binom{\beta}{2}}x^{h-2}S^{(2,k)}(x)
		%	+(1+x^{k-1})(-1)^{\binom{\beta+1}{2}}x^{h-1}S^{(1,k)}(x)=0.
		%\end{equation*}
		%
		%By simplifying the term $x^{h-2}$, we have to prove that:
		%\begin{equation*}
		%	(-1)^{\binom{\beta-1}{2}} (x^2-3x-x^{k+1}+1)-(-1)^{\binom{\beta}{2}} 
		%	(2x-x^k-1)+ (-1)^{\binom{\beta+1}{2}} (x^2+x^{k+1}-x-x^k)=0.
		%\end{equation*}
		%It is easy to see that $\binom{\beta-1}{2}$ and $\binom{\beta+1}{2}$ always 
		%have different parity. 
		%Since $\beta$ is even, $\binom{\beta-1}{2}$ and $\binom{\beta}{2}$ also have 
		%different parity. 
		%Therefore, we have either
		%\[
		%\binom{\beta-1}{2}\ \text{even},\quad
		%\binom{\beta}{2},\ \binom{\beta+1}{2}\ \text{odd},
		%\]
		%or
		%\[
		%\binom{\beta-1}{2}\ \text{odd},\quad
		%\binom{\beta}{2},\ \binom{\beta+1}{2}\ \text{even}.
		%\]
		%
		%In both the cases, the thesis follows.
		\item Concerning the $(h-2)$-th equation, after performing the 
		substitutions, the following relation must be satisfied: 
		\begin{align*}
			&(-1)^{\binom{\beta-2}{2}} x^{h-3} S^{(4,k)}(x)-(-1)^{\binom{\beta-1}{2}}x^{h-3}S^{(3,k)}(x)+\\
			&(-1)^{\binom{\beta}{2}}x^{h-2}S^{(2,k)}(x)-x^{k-1}(-1))^{\binom{\beta+1}{2}} x^{h-1}S^{(2,k)}(x)=\\
			&(-1)^{\binom{\beta-2}{2}} (3x^{h-1}-4x^{h-2}+x^{k+h-2}-x^{k+h-1}+x^{h-3})-\\
			&(-1)^{\binom{\beta-1}{2}}(x^{h-1}-3x^{h-2}+x^{k+h-2}+x^{h-3})+\\
			&(-1)^{\binom{\beta}{2}}(2x^{h-1}-x^{k+h-2}-x^{h-2})-\\
			&(-1)^{\binom{\beta+1}{2}} (x^{k+h-1}-x^{k+h-2})=0\ .
		\end{align*}
		
		It is easy to prove that 
		$\binom{\beta-2}{2}$ and $\binom{\beta}{2}$ have different parity, as well as 
		$\binom{\beta-1}{2}$ and $\binom{\beta+1}{2}$. Therefore, we have either
		\[
		\binom{\beta-2}{2}\mbox{ even }, \binom{\beta-1}{2}\ \text{even},
		\binom{\beta}{2}\mbox{ odd },\ \binom{\beta+1}{2}\ \text{odd},
		\]
		or
		\[
		\binom{\beta-2}{2}\mbox{ odd }, \binom{\beta-1}{2}\ \text{odd},
		\binom{\beta}{2}\mbox{ even },\ \binom{\beta+1}{2}\ \text{even}.
		\]
		In either case, the thesis holds.
		
		%\item If $i=3$, then we have to prove 
		%$xF_{h-3}(x)-F_{h-2}(x)+F_{h-1}-x^{k-1}F_h(x)=0$. Posing $\beta=(h \bmod 2 + 
		%h + 2)$ and applying expressions in (\ref{candidate}), we have to prove that:
		%	\begin{align*}
			%		&(-1)^{\binom{\beta-2}{2}} x^{h-3} 
			%		S^{(4,k)}(x)-(-1)^{\binom{\beta-1}{2}}x^{h-3}S^{(3,k)}(x)+\\
			%		&(-1)^{\binom{\beta}{2}}x^{h-2}S^{(2,k)}(x)-x^{k-1}(-1))^{\binom{\beta
					%		+1}{2}} x^{h-1}S^{(2,k)}(x)=\\
			%		&(-1)^{\binom{\beta-2}{2}} 
			%		(3x^{h-1}-4x^{h-2}+x^{k+h-2}-x^{k+h-1}+x^{h-3})-\\
			%		&(-1)^{\binom{\beta-1}{2}}(x^{h-1}-3x^{h-2}+x^{k+h-2}+x^{h-3})+\\
			%		&(-1)^{\binom{\beta}{2}}(2x^{h-1}-x^{k+h-2}-x^{h-2})-\\
			%		&(-1)^{\binom{\beta+1}{2}} (x^{k+h-1}-x^{k+h-2})=0.
			%	\end{align*}
		%	
		%	We know that $\binom{\beta-1}{2}$ and $\binom{\beta+1}{2}$, also 
		%	$\binom{\beta-2}{2}$ and $\binom{\beta}{2}$, always have different 
		%	parity. Since $\beta$ is even, $\binom{\beta-2}{2}$ and 
		%	$\binom{\beta-1}{2}$ have same parity. 
		%	Therefore, we have either
		%	\[
		%	\binom{\beta-2}{2} \binom{\beta-1}{2}\ \text{even},\quad
		%	\binom{\beta}{2},\ \binom{\beta+1}{2}\ \text{odd},
		%	\]
		%	or
		%	\[
		%	\binom{\beta-2}{2} \binom{\beta-1}{2}\ \text{odd},\quad
		%	\binom{\beta}{2},\ \binom{\beta+1}{2}\ \text{even}.
		%	\]
		%
		%In both the cases, the thesis follows.
		\item Verifying one of the equations of (\ref{F2_hk}) from the $(h-3)$-th to the second equation, i.e.,  
		$$x F_{h-i}(x) - F_{h-i+1}(x) + F_{h-i+2}(x)=0\ \mbox{ for } i=4,5,\ldots,h-1\ , $$
		after the usual substitutions and setting 
		$$
		\alpha=(h \bmod 2 + h - i + 3)\ ,
		$$
		is equivalent to verifying that   
		\begin{equation*}
			x(-1)^{\binom{\alpha}{2}} x^{h-i-1} S^{(i+1,k)}(x) -(-1)^{\binom{\alpha+1}{2}} x^{h-i} S^{(i,k)}(x)+(-1)^{\binom{\alpha+2}{2}} x^{h-i+1}S^{(i-1,k)}(x)=0\ .	
		\end{equation*}
		%\item For each $i=4,5,\ldots,h-1$, we prove that $x F_{h-i}(x) - F_{h-i+1}(x) 
		%+ F_{h-i+2}(x)=0$.
		%
		%Fixed $i$ and posing $\alpha=(h \bmod 2 + h - i + 3)$, we have to prove that 
		%\begin{equation*}
		%x(-1)^{\binom{\alpha}{2}} x^{h-i-1} S^{(i+1,k)}(x) 
		%-(-1)^{\binom{\alpha+1}{2}} x^{h-i} S^{(i,k)}(x)+(-1)^{\binom{\alpha+2}{2}} 
		%x^{h-i+1}S^{(i-1,k)}(x)=0	
		%\end{equation*}
		%by applying the expressions in (\ref{candidate}).
		
		By simplifying the term $x^{h-i}$ and using expressions in (\ref{S}), the equation to prove boils down to:
		\begin{align*}
			&(-1)^{\binom{\alpha}{2}} \Bigg( \sum_{j=0}^{\left\lfloor \frac{i+2}{2} \right\rfloor}
			(-1)^{\binom{i+2}{2}-j}
			\binom{i-j+2}{j}
			x^j+
			\sum_{j=1}^{\left\lfloor \frac{i+1}{2} \right\rfloor}
			(-1)^{\binom{i+2}{2}-j+1}
			\binom{i-j}{j-1}
			x^{k+j} \Bigg)-\\
			&\\
			&(-1)^{\binom{\alpha+1}{2}} \Bigg( \sum_{j=0}^{\left\lfloor \frac{i+1}{2} \right\rfloor}
			(-1)^{\binom{i+1}{2}-j}
			\binom{i-j+1}{j}
			x^j+
			\sum_{j=1}^{\left\lfloor \frac{i}{2} \right\rfloor}
			(-1)^{\binom{i+1}{2}-j+1}
			\binom{i-j-1}{j-1}
			x^{k+j} \Bigg)+\\
			&\\
			&(-1)^{\binom{\alpha+2}{2}} \Bigg( \sum_{j=0}^{\left\lfloor \frac{i}{2} \right\rfloor}
			(-1)^{\binom{i}{2}-j}
			\binom{i-j}{j}
			x^{j+1}+
			\sum_{j=1}^{\left\lfloor \frac{i-1}{2} \right\rfloor}
			(-1)^{\binom{i}{2}-j+1}
			\binom{i-j-2}{j-1}
			x^{k+j+1} \Bigg)=0.
		\end{align*}
		After reindexing the last two sums, the expression reduces to
		\begin{align}\label{EXPR}
			&(-1)^{\binom{\alpha}{2}} \Bigg( \sum_{j=0}^{\left\lfloor \frac{i+2}{2} \right\rfloor}
			(-1)^{\binom{i+2}{2}-j}
			\binom{i-j+2}{j}
			x^j+
			\sum_{j=1}^{\left\lfloor \frac{i+1}{2} \right\rfloor}
			(-1)^{\binom{i+2}{2}-j+1}
			\binom{i-j}{j-1}
			x^{k+j} \Bigg)-\\ \nonumber
			&\\\nonumber
			&(-1)^{\binom{\alpha+1}{2}} \Bigg( \sum_{j=0}^{\left\lfloor \frac{i+1}{2} \right\rfloor}
			(-1)^{\binom{i+1}{2}-j}
			\binom{i-j+1}{j}
			x^j+
			\sum_{j=1}^{\left\lfloor \frac{i}{2} \right\rfloor}
			(-1)^{\binom{i+1}{2}-j+1}
			\binom{i-j-1}{j-1}
			x^{k+j} \Bigg)+\\\nonumber
			&\\\nonumber
			&(-1)^{\binom{\alpha+2}{2}} \Bigg( \sum_{j=1}^{\left\lfloor \frac{i+2}{2} \right\rfloor}
			(-1)^{\binom{i}{2}-j+1}
			\binom{i-j+1}{j-1}
			x^{j}+
			\sum_{j=2}^{\left\lfloor \frac{i+1}{2} \right\rfloor}
			(-1)^{\binom{i}{2}-j+2}
			\binom{i-j-1}{j-2}
			x^{k+j} \Bigg)=0.
		\end{align}
		We precede by showing that the coefficients of the powers of $x$ are zero.
		\begin{itemize}
			\item The coefficient of $x^0$ and $x^{k+1}$ is 
			\[
			(-1)^{\binom{\alpha}{2}+{\binom{i+2}{2}}}- 
			(-1)^{\binom{\alpha+1}{2}+{\binom{i+1}{2}}}
			\]
			and it is zero depending on the parity of $h$, having set $\alpha=h 
			\bmod 2 + h - i + 3$. More precisely,
			if $h$ is even, we have
			$(-1)^{\binom{h-i+3}{2}+\binom{i+2}{2}}- 
			(-1)^{\binom{h-i+4}{2}+\binom{i+1}{2}}=0$ if and only if the exponents 
			have the same parity.
			It is not difficult to show that the two exponents differ by an even 
			integer, which implies that their parities are equal. 
			Therefore, both terms cancel each other, resulting in the expression 
			being zero.
			
			If $h$ is odd, then we have
			$(-1)^{\binom{h-i+4}{2}+\binom{i+2}{2}}- 
			(-1)^{\binom{h-i+5}{2}+\binom{i+1}{2}}=0$ if and only if the exponents 
			have the same parity.
			Also in this case, the two exponents differ by an even integer, so that their parities are equal. Therefore, both terms cancel each 
			other, resulting in the expression being zero.
			
			\noindent
			From the above, the expression (\ref{EXPR}) to be proved becomes:
			{\small
				\begin{align}\label{EXPR2}
					&(-1)^{\binom{\alpha}{2}} \Bigg( \sum_{j=1}^{\left\lfloor \frac{i+2}{2} \right\rfloor}
					(-1)^{\binom{i+2}{2}-j}
					\binom{i-j+2}{j}
					x^j+
					\sum_{j=2}^{\left\lfloor \frac{i+1}{2} \right\rfloor}
					(-1)^{\binom{i+2}{2}-j+1}
					\binom{i-j}{j-1}
					x^{k+j} \Bigg)-\\ \nonumber
					%	&\\\nonumber
					&(-1)^{\binom{\alpha+1}{2}} \Bigg( \sum_{j=1}^{\left\lfloor \frac{i+1}{2} \right\rfloor}
					(-1)^{\binom{i+1}{2}-j}
					\binom{i-j+1}{j}
					x^j+
					\sum_{j=2}^{\left\lfloor \frac{i}{2} \right\rfloor}
					(-1)^{\binom{i+1}{2}-j+1}
					\binom{i-j-1}{j-1}
					x^{k+j} \Bigg)+\\\nonumber
					%	&\\\nonumber
					&(-1)^{\binom{\alpha+2}{2}} \Bigg( \sum_{j=1}^{\left\lfloor \frac{i+2}{2} \right\rfloor}
					(-1)^{\binom{i}{2}-j+1}
					\binom{i-j+1}{j-1}
					x^{j}+
					\sum_{j=2}^{\left\lfloor \frac{i+1}{2} \right\rfloor}
					(-1)^{\binom{i}{2}-j+2}
					\binom{i-j-1}{j-2}
					x^{k+j} \Bigg)=0
				\end{align}
			}
			(observe that the lower limits of the sums for \(x^j\) and \(x^{k+j}\) are the same in all three terms).
			\item We now analyse the coefficient of $x^j$ in (\ref{EXPR2}). First, we assume that $i$ is odd. Then
			\[
			\left\lfloor \frac{i+2}{2} \right\rfloor = \left\lfloor \frac{i+1}{2} \right\rfloor.
			\]
			Consequently, for $j = 1, 2, \ldots, \left\lfloor \frac{i+1}{2} \right\rfloor$, the coefficient is given by
			\begin{align*}
				&(-1)^{\binom{\alpha}{2}}(-1)^{\binom{i+2}{2}-j}\binom{i-j+2}{j}
				- (-1)^{\binom{\alpha+1}{2}}(-1)^{\binom{i+1}{2}-j}\binom{i-j+1}{j} \\
				&+(-1)^{\binom{\alpha+2}{2}}(-1)^{\binom{i}{2}-j+1}\binom{i-j+1}{j-1}= \\
				& \underbrace{(-1)^{\binom{\alpha}{2}+\binom{i+2}{2}-j}\binom{i-j+1}{j-1}}_{A}
				+ \underbrace{(-1)^{\binom{\alpha}{2}+\binom{i+2}{2}-j}\binom{i-j+1}{j}}_{B} \\
				&- \underbrace{(-1)^{\binom{\alpha+1}{2}+\binom{i+1}{2}-j}\binom{i-j+1}{j}}_{C}
				+ \underbrace{(-1)^{\binom{\alpha+2}{2}+\binom{i}{2}-j+1}\binom{i-j+1}{j-1}}_{D}
			\end{align*}
			(here, we have used the Pascal's formula about binomial coefficients in the first term).
			
			Such an expression is equal to $0$, since it is routine prove that the exponents in $A$ and $D$ have opposite parity, whereas the ones in $B$ and $C$ have the same parity.
			
			\bigskip
			If $i$ is even, then
			\[
			\left\lfloor \frac{i+1}{2} \right\rfloor = \left\lfloor \frac{i+2}{2} \right\rfloor - 1.
			\]
			Consequently, for $j = 1, 2, \ldots, \left\lfloor \frac{i+2}{2} \right\rfloor - 1$, the coefficient of $x^j$ is the same as in the case when $i$ is odd, hence it is zero by a similar argument. While, for $j=\left\lfloor \frac{i+2}{2} \right\rfloor$, the coefficient of $x^j$ is
			\[
			(-1)^{\binom{\alpha}{2}+\binom{i+2}{2}-\left\lfloor \frac{i+2}{2} \right\rfloor}\binom{i-\left\lfloor \frac{i+2}{2} \right\rfloor+2}{\left\lfloor \frac{i+2}{2} \right\rfloor}
			+ (-1)^{\binom{\alpha+2}{2}+\binom{i}{2}-\left\lfloor \frac{i+2}{2} \right\rfloor+1}\binom{i-\left\lfloor \frac{i+2}{2} \right\rfloor+1}{\left\lfloor \frac{i+2}{2} \right\rfloor-1}.
			\]
			Since $i$ is even, it reduces to: 
			\[
			(-1)^{\binom{\alpha}{2}+\binom{i+2}{2}-\left\lfloor \frac{i+2}{2} \right\rfloor}
			+ (-1)^{\binom{\alpha+2}{2}+\binom{i}{2}-\left\lfloor \frac{i+2}{2} \right\rfloor+1}.
			\]
			Moreover, the terms $\binom{\alpha}{2}$ and $\binom{\alpha+2}{2}$ have opposite parity, as do $\binom{i+2}{2}$ and $\binom{i}{2}$. Therefore, the two exponents of $-1$ differ by an odd integer and hence have opposite parity. Thus, the sum is zero.
			\item We now analyse the coefficient of $x^{k+j}$ in (\ref{EXPR2}). First, we assume that $i$ is even. Then
			\[
			\left\lfloor \frac{i+1}{2} \right\rfloor = \left\lfloor \frac{i}{2} \right\rfloor.
			\]
			Consequently, for $j=2,3,\ldots,\left\lfloor \frac{i}{2} \right\rfloor$, the coefficient of $x^{k+j}$ is:
			
			\begin{align*}
				&(-1)^{\binom{\alpha}{2}}(-1)^{\binom{i+2}{2}-j+1}\binom{i-j}{j-1}
				- (-1)^{\binom{\alpha+1}{2}}(-1)^{\binom{i+1}{2}-j+1}\binom{i-j-1}{j-1} \\
				&+(-1)^{\binom{\alpha+2}{2}}(-1)^{\binom{i}{2}-j+2}\binom{i-j-1}{j-2}= \\
				& \underbrace{(-1)^{\binom{\alpha}{2}+\binom{i+2}{2}-j+1}\binom{i-j-1}{j-2}}_{A}
				+ \underbrace{(-1)^{\binom{\alpha}{2}+\binom{i+2}{2}-j+1}\binom{i-j-1}{j-1}}_{B} \\
				&- \underbrace{(-1)^{\binom{\alpha+1}{2}+\binom{i+1}{2}-j+1}\binom{i-j-1}{j-1}}_{C}
				+ \underbrace{(-1)^{\binom{\alpha+2}{2}+\binom{i}{2}-j+2}\binom{i-j-1}{j-2}}_{D}
			\end{align*}
			(also here, we have used the Pascal's formula about binomial coefficients in the first term).
			
			Also in this case such an expression is equal to $0$, since it is routine prove that the exponents in $A$ and $D$ have opposite parity, whereas the ones in $B$ and $C$ have the same parity.
			
			\bigskip
			If $i$ is odd, then
			\[
			\left\lfloor \frac{i+1}{2} \right\rfloor = \left\lfloor \frac{i}{2} \right\rfloor + 1.
			\]
			Therefore, for $j=1,2,\ldots,\left\lfloor \frac{i}{2} \right\rfloor$ the coefficient of $x^{k+j}$ is the same as in the case when $i$ is even, hence it is zero by a similar argument. While, for $j=\left\lfloor \frac{i}{2} \right\rfloor + 1$, the coefficient of $x^{k+j}$ is:
			\[
			(-1)^{\binom{\alpha}{2}+\binom{i+2}{2}-\left\lfloor \frac{i}{2} \right\rfloor}\binom{i-\left\lfloor \frac{i}{2} \right\rfloor-1}{\left\lfloor \frac{i}{2} \right\rfloor}
			+ (-1)^{\binom{\alpha+2}{2}+\binom{i}{2}-\left\lfloor \frac{i}{2} \right\rfloor+1}\binom{i-\left\lfloor \frac{i}{2} \right\rfloor-2}{\left\lfloor \frac{i}{2} \right\rfloor-1}.
			\]
			Since $i$ is odd, it reduces to: 
			\[
			(-1)^{\binom{\alpha}{2}+\binom{i+2}{2}-\left\lfloor \frac{i}{2} \right\rfloor}
			+ (-1)^{\binom{\alpha+2}{2}+\binom{i}{2}-\left\lfloor \frac{i}{2} \right\rfloor+1}.
			\]
			Moreover, the terms $\binom{\alpha}{2}$ and $\binom{\alpha+2}{2}$ have opposite parity, as do $\binom{i+2}{2}$ and $\binom{i}{2}$. Therefore, the two exponents of $-1$ differ by an odd integer and hence have opposite parity. Thus, the sum is zero.
			
		\end{itemize}
	\end{itemize}
	
	The proof is completed.

	\section{Combinatorial identity involving Catalan numbers}\label{catalan}
	
	The generating function of the paths in $\mathcal D^{(h,k)}$ according to their semilength is
	
	\begin{align*}
		&f(x)=D^{(h,k)}(x)=\sum_{n \geq 0} D_n^{(h,k)} x^n= \sum_{\ell=1}^{h} F_\ell(x)
		+ \frac{x^{k-1} - x}{x - 1} F_h^{(h,k)}(x)=\\
		&\sum_{\ell=1}^{h} (-1)^{\binom{h \bmod 2+ \ell+3}{2}} x^{\ell-1}\frac {S^{(h+1-\ell,k)}(x)}{S^{(h,k)}(x)} + \frac{x^{k-1} - x}{x - 1}(-1)^{\binom{h \bmod 2+ h+3}{2}} x^{h-1}\frac {S^{(1,k)}(x)}{S^{(h,k)}(x)}.
	\end{align*}
	
	Thus,
	
	\[
	S^{(h,k)}(x) D^{h,k}(x) = \sum_{\ell=1}^{h} (-1)^{\binom{h \bmod 2+ \ell+3}{2}} x^{\ell-1} S^{h+1-\ell,k}(x) + x^h \left( x^{k-2} - 1 \right)(-1)^{\binom{h \bmod 2+ h+3}{2}}
	\]
	
	\noindent
	Then, from expression in (\ref{S}), we have:
	{\small
		\begin{align}\label{PIPPO}
			\nonumber&\left(\sum_{j=0}^{\lfloor \frac{h+1}{2} \rfloor} (-1)^{\binom{h+1}{2}-j} \binom{h+1-j}{j}x^{j} + \sum_{j=1}^{\lfloor \frac{h}{2} \rfloor} (-1)^{\binom{h+1}{2}-j+1} \binom{h-j+1}{j-1} x^{k+j}\right) \left( \sum_{n\ge 0} D^{(h,k)}_n x^n \right) =\\
			\nonumber&\\
			&\left(\sum_{\ell=1}^{h} \sum_{j=0}^{\lfloor \frac{h+2-\ell}{2} \rfloor} (-1)^{\binom{h \bmod 2 + \ell + 3}{2} + \binom{h+2-\ell}{2}-j} \binom{h+2-\ell-j}{j} x^{\ell-1+j}\right) +\\ \nonumber
			\nonumber&\\
			&
			\nonumber\left(\sum_{\ell=1}^{h} \sum_{j=1}^{\lfloor \frac{h+1-\ell}{2} \rfloor} (-1)^{\binom{h \bmod 2 + \ell + 3}{2} + \binom{h+2-\ell}{2}-j+1} \binom{h-\ell-j}{j-1} x^{k+\ell-1+j}\right) + x^h (x^{k-2}-1) (-1)^{\binom{h \bmod 2+h+3}{2}}.
		\end{align}
	}
	
	\noindent
	The above equation can be reformulated to explicitly display the coefficient of each power of $x$. To this purpose, recalling that
	\[
	\left(\displaystyle\sum_{j=0}^tb_jx^j\right)
	\left(\displaystyle\sum_{n\geq 0}a_nx^n\right)
	=
	\displaystyle\sum_{n\geq 0}
	\left(\displaystyle\sum_{j=0}^ta_{n-j}b_j\right)
	x^n
	\]
	where $a_{n-j}=0 \mbox{ if } n-j<0$, then the left-hand side of equation (\ref{PIPPO}) can be written as
	
	{\small
		\begin{equation*}
			\sum_{n\geq 0}\left(\sum_{j=0}^{\lfloor \frac{h+1}{2} \rfloor} D_{n-j}^{(h,k)}(-1)^{\binom{h+1}{2}-j} \binom{h+1-j}{j}\right)x^n
			+
			x^k\sum_{n\geq 0}\left(\sum_{j=1}^{\lfloor \frac{h}{2} \rfloor} D_{n-j}^{(h,k)}(-1)^{\binom{h+1}{2}-j+1} \binom{h+1-j}{j-1}\right)x^n\ ,
		\end{equation*}
		where $D_{n-j}^{(h,k)}=0\text{ if } n-j<0$.
	}
	
	\noindent
	The first term of the right-hand side of equation (\ref{PIPPO}), namely
	\begin{equation}\label{secondMember}
		\sum_{\ell=1}^{h} \sum_{j=0}^{\lfloor \frac{h+2-\ell}{2} \rfloor} (-1)^{\binom{h \bmod 2 + \ell + 3}{2} + \binom{h+2-\ell}{2}-j} \binom{h+2-\ell-j}{j} x^{\ell-1+j}\ ,
	\end{equation}
	can be replaced by
	
	\begin{equation}\label{secondMember2}
		\sum_{\ell=1}^{h} \sum_{j=0}^{\lfloor \frac{h+2-\ell}{2} \rfloor} (-1)^{\lfloor \frac{h+1}{2} \rfloor-j} \binom{h+2-\ell-j}{j} x^{\ell-1+j}\ ,
	\end{equation}
	since it is easily to show that  $\binom{h \bmod 2 + \ell + 3}{2} + \binom{h+2-\ell}{2}$ has the same parity of \(\left\lfloor\frac{h+1}{2}\right\rfloor\).
	
	In order to highlight the powers of $x$ and their corresponding coefficients, as before, the substitutions $\ell = n+1-t$ and $j = t$ are performed. 
	To determine the upper and lower bounds of the new variables $n$ and $t$, we observe the following.
	
	\begin{itemize}
		\item Since $\ell \geq 1$, $j \geq 0$, and $n = \ell - 1 + j = \ell - 1 + t \geq 0$, it follows that $n \geq 0$. 
		The upper bound of $n$, which is $n=h$, is attained by setting $\ell = h$, which then implies $j = 1$, in the expression $n = \ell - 1 + j=\ell - 1 + t$. Therefore,
		\[
		0\leq n\leq h\ .
		\]
		\item For the bounds of the index $t$ we note:
		\begin{itemize}
			\item from $\ell\geq 1$ in (\ref{secondMember}) and  $t=n-\ell+1$, we have $t\leq n$;
			\item from $j\leq \left\lfloor \frac{h+2-\ell}{2} \right\rfloor$ and $\ell=n-t+1$, we obtain $j\leq \left\lfloor \frac{h+1-n+t}{2} \right\rfloor$ equivalent to $t\leq \left\lfloor \frac{h+1-n+t}{2} \right\rfloor$, so $t\leq h-n+1$;
			\item from $\ell\leq h$ and $j\geq 0$ in (\ref{secondMember}) we deduce $t\geq n-h+1$ and $t\geq 0$, respectively. Therefore,
			\[
			\max\{0,n-h+1\}\leq t \leq \min\{n,h-n+1\}\ .
			\]
			Since $n=0,1,2,\ldots,h$, then $\max\{0,n-h+1\}=\left\lfloor \frac{n}{h} \right\rfloor$.
		\end{itemize}
	\end{itemize}

	Summarizing, equation (\ref{secondMember2}) can be reformulated as
	\begin{equation}\label{secMemberBis}
		\sum_{n=0}^h\sum_{t={\left\lfloor \frac{n}{h} \right\rfloor}}^{\min\{n, h-n+1\}} (-1)^{\left\lfloor \frac{h+1}{2} \right\rfloor-t} \binom{h-n+1}{t}x^n \ .
	\end{equation}
	
	Similarly, the second term of the right-end side of equation (\ref{PIPPO})
	becomes:
	
	\begin{equation*}
		x^k\sum_{n=1}^{h-1}\sum_{t=n}^{\min\{n, h-n\}} (-1)^{\left\lfloor \frac{h+1}{2} \right\rfloor-t+1} \binom{h-n-1}{t-1}x^n\ .
	\end{equation*}
	Therefore, equation (\ref{PIPPO}) is:

	{\footnotesize
		\begin{align}\label{PIPPO2}
			\nonumber&\sum_{n\geq 0}\left(\sum_{j=0}^{\left\lfloor \frac{h+1}{2} \right\rfloor} D_{n-j}^{(h,k)}(-1)^{\binom{h+1}{2}-j} \binom{h+1-j}{j}\right)x^n
			+
			x^k\sum_{n\geq 0}\left(\sum_{j=1}^{\lfloor \frac{h}{2} \rfloor} D_{n-j}^{(h,k)}(-1)^{\binom{h+1}{2}-j+1} \binom{h+1-j}{j-1}\right)x^n =\\ \nonumber
			&\\
			&\sum_{n=0}^h\left(\sum_{t={\left\lfloor \frac{n}{h} \right\rfloor}}^{\min\{n, h-n+1\}} (-1)^{\left\lfloor\frac{h+1}{2}\right\rfloor-t} \binom{h-n+1}{t}\right)x^n + \\ \nonumber
			&\\
			\nonumber&x^k\sum_{n=1}^{h-1}\left(\sum_{t=n}^{\min\{n, h-n\}} (-1)^{\left\lfloor\frac{h+1}{2}\right\rfloor-t+1} \binom{h-n-1}{t-1}\right)x^n + x^h (x^{k-2}-1) (-1)^{\binom{h \bmod 2+h+3}{2}}.
		\end{align}
	}
	
	\noindent
	This new version (\ref{PIPPO2}) of equation (\ref{PIPPO}) easily allows to see that, if $n<h<k$, the following relation between the coefficients of $x^n$ in the left-hand and right hand side holds: 
	\begin{equation}\label{RELAZIONE}
		\sum_{j=0}^{\left\lfloor \frac{h+1}{2} \right\rfloor} D_{n-j}^{(h,k)}(-1)^{\binom{h+1}{2}-j} \binom{h+1-j}{j}
		=
		\sum_{t={\left\lfloor \frac{n}{h} \right\rfloor}}^{\min\{n, h-n+1\}} (-1)^{\left\lfloor\frac{h+1}{2}\right\rfloor-t} \binom{h-n+1}{t}\ .
	\end{equation}

	When $n \leq h$, the set $D_{n}^{(h,k)}$ corresponds to the set of the unrestricted Dyck paths having semilength $n \geq 0$ which are enumerated by the $n$-Catalan number, that is $C_n$. In our case, we have $n \lneq h$, so $\left\lfloor \frac{n}{h} \right\rfloor=0$ and equation (\ref{RELAZIONE}) becomes: 
	
	\begin{equation}\label{RELAZIONE2}
		\sum_{j=0}^{\left\lfloor \frac{h+1}{2} \right\rfloor} C_{n-j}^{(h,k)}(-1)^{\binom{h+1}{2}-j} \binom{h+1-j}{j}
		=
		\sum_{t=0}^{\min\{n, h-n+1\}} (-1)^{\left\lfloor\frac{h+1}{2}\right\rfloor-t} \binom{h-n+1}{t}\ .
	\end{equation}
	
	Moreover, if $\min\{n, h-n+1\}=h-n+1$, equivalently if $n \geq \left\lceil \frac{h+1}{2} \right\rceil$, then the right-hand side of (\ref{RELAZIONE2}) vanishes, since it reduces to the alternating sum of the entries in a row of Pascal's triangle. Definitely, if $\left\lceil \frac{h+1}{2} \right\rceil \leq n < h$, we have:
	\begin{equation*}
		\sum_{j=0}^{\left\lfloor \frac{h+1}{2} \right\rfloor} C_{n-j}^{(h,k)}(-1)^{\binom{h+1}{2}-j} \binom{h+1-j}{j}
		=0 \,
	\end{equation*}
	which boils down to: 
	\begin{equation}\label{RELAZIONE3}
		\sum_{j=0}^{\left\lfloor \frac{h+1}{2} \right\rfloor} C_{n-j}^{(h,k)}(-1)^{j} \binom{h+1-j}{j}
		=0 \, .
	\end{equation}
	Equation (\ref{RELAZIONE3}) allows a new intriguing relation for Catalan number defined as:
	\begin{equation*}
		C_n=\sum_{j=1}^{\left\lfloor \frac{h+1}{2} \right\rfloor}  (-1)^{j+1} \binom{h+1-j}{j}C_{n-j} \, \text{ for } \left\lceil \frac{h+1}{2} \right\rceil \leq n < h \, .
	\end{equation*}
	
	\section{Conclusions}
	
	In this paper we investigated Dyck paths of height at most $h$ with the additional restriction of avoiding $k-1$ consecutive valleys at height $h-1$. By combining classical techniques for the enumeration of constrained lattice paths, we obtained the generating function of the corresponding class of paths.
	
	As a direct consequence of this analysis, we derived a new identity involving Catalan numbers. To the best of our knowledge, this relation does not appear in the existing literature. In contrast with previously known identities arising from similar restrictions, the recurrence relation obtained for the $n$-th Catalan number is characterized, under suitable hypotheses, by constant coefficients. We note that the classical Catalan recurrences have coefficients depending on $n$. Probably, the occurrence of constant coefficients highlights a possible structural regularity induced by the combined height and valley constraints.
	
	The results presented here suggest further developments. A natural open problem is to replace the restriction on valleys at height $h-1$ with the avoidance of $k-1$ consecutive valleys at a generic height $i < h$. Such a generalization would likely produce new recurrence relations for Catalan numbers.
	
	We believe that the approach introduced in this work can serve as a flexible framework for studying additional local constraints on Dyck paths and for uncovering new relations within the rich combinatorial theory of Catalan numbers.

\end{document}